\documentclass{vgtc}                        
\graphicspath{{figures/}{pictures/}{images/}{./}} 
\usepackage{times}

\usepackage{tabu}                      
\usepackage{booktabs}                 
\usepackage{lipsum}                    
\usepackage{mwe}                      
\usepackage{graphicx} 
\usepackage{amsmath} \usepackage{amssymb}

\usepackage{mathptmx}  
\newcommand{\revision}[1]{\textcolor{black}{#1}}

\onlineid{1064}
\vgtccategory{Research}
\vgtcinsertpkg
\usepackage{makecell}
\usepackage{arydshln}
\usepackage{latexsym}
\usepackage{multirow}
\usepackage{enumitem}
\usepackage{float}
\usepackage{wasysym}

\usepackage{xcolor}
\definecolor{b}{HTML}{1d91c0}
\definecolor{o}{HTML}{fdae6b}
\definecolor{emphasize}{HTML}{4874CB}

\let\oldcolorbox\colorbox

\renewcommand{\colorbox}[2]{
  \setlength{\fboxsep}{0pt} 
  \oldcolorbox{#1}{
    \raisebox{0pt}[8pt][3pt]{\strut #2}
  }
}

\newcommand{\squarecolorbox}[2]{
  \setlength{\fboxsep}{0pt} 
  \setlength{\fboxrule}{0pt} 
  \colorbox{#1}{
    \makebox[0.8em][c]{\hfill\textcolor{white}{\small\strut #2}\hfill}
  }
}

\usepackage{longtable}

\usepackage{booktabs}
\usepackage{caption}
\begin{document}

\title{A Comparative Study of Table-Sized Physicalization and Digital Visualization}

\author{Yanxin Wang\thanks{e-mail: yanxin.wang20@student.xjtlu.edu.cn} %
\and Yihan Liu\thanks{e-mail:yihan.liu17@student.xjtlu.edu.cn} %
\and Lingyun Yu\thanks{e-mail:Lingyun.Yu@xjtlu.edu.cn}
\and Chengtao Ji\thanks{e-mail:Chengtao.Ji@xjtlu.edu.cn}
\and Yu Liu\thanks{e-mail:Yu.Liu02@xjtlu.edu.cn}
}

\affiliation{\scriptsize School of Advanced Technology \\ Xi'an Jiaotong-Liverpool University}

\teaser{
\vspace{-4mm}
  \centering
  \includegraphics[width=\linewidth]{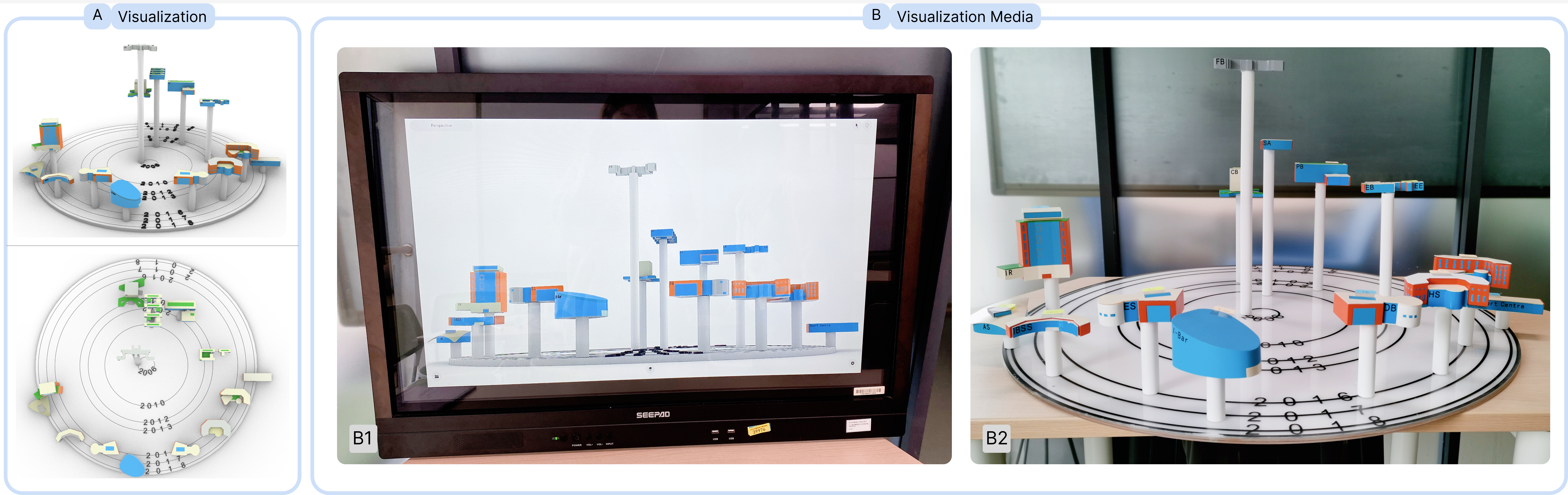}
  \caption{\textbf{A} Multiple perspectives of our visualization; \textbf{B} Two mediums of visualization: \textbf{B1} Screen-Based Digital Visualization \textbf{B2} Physicalization}
  \label{fig:teaser}
}

\abstract{
Data physicalization is gaining popularity in public and educational contexts due to its potential to make abstract data more tangible and understandable. Despite its growing use, there remains a significant gap in our understanding of how large-size physical visualizations compare to their digital counterparts in terms of user comprehension and memory retention. This study aims to bridge this knowledge gap by comparing the effectiveness of visualizing school building history data on large digital screens versus large physical models.
Our experimental approach involved 32 participants who were exposed to one of the visualization mediums. We assessed their user experience and immediate understanding of the content, measured through tests after exposure, and evaluated memory retention with follow-up tests seven days later. The results revealed notable differences between the two forms of visualization: physicalization not only facilitated better initial comprehension but also significantly enhanced long-term memory retention. Furthermore, user feedback on usability was also higher on physicalization.
These findings underscore the substantial impact of physicalization in improving information comprehension and retention. This study contributes crucial insights into future visualization media selection in educational and public settings.

} 

\firstsection{Introduction}

\maketitle
\begin{figure*}[htb]
\centering
\includegraphics[width=0.98\textwidth]{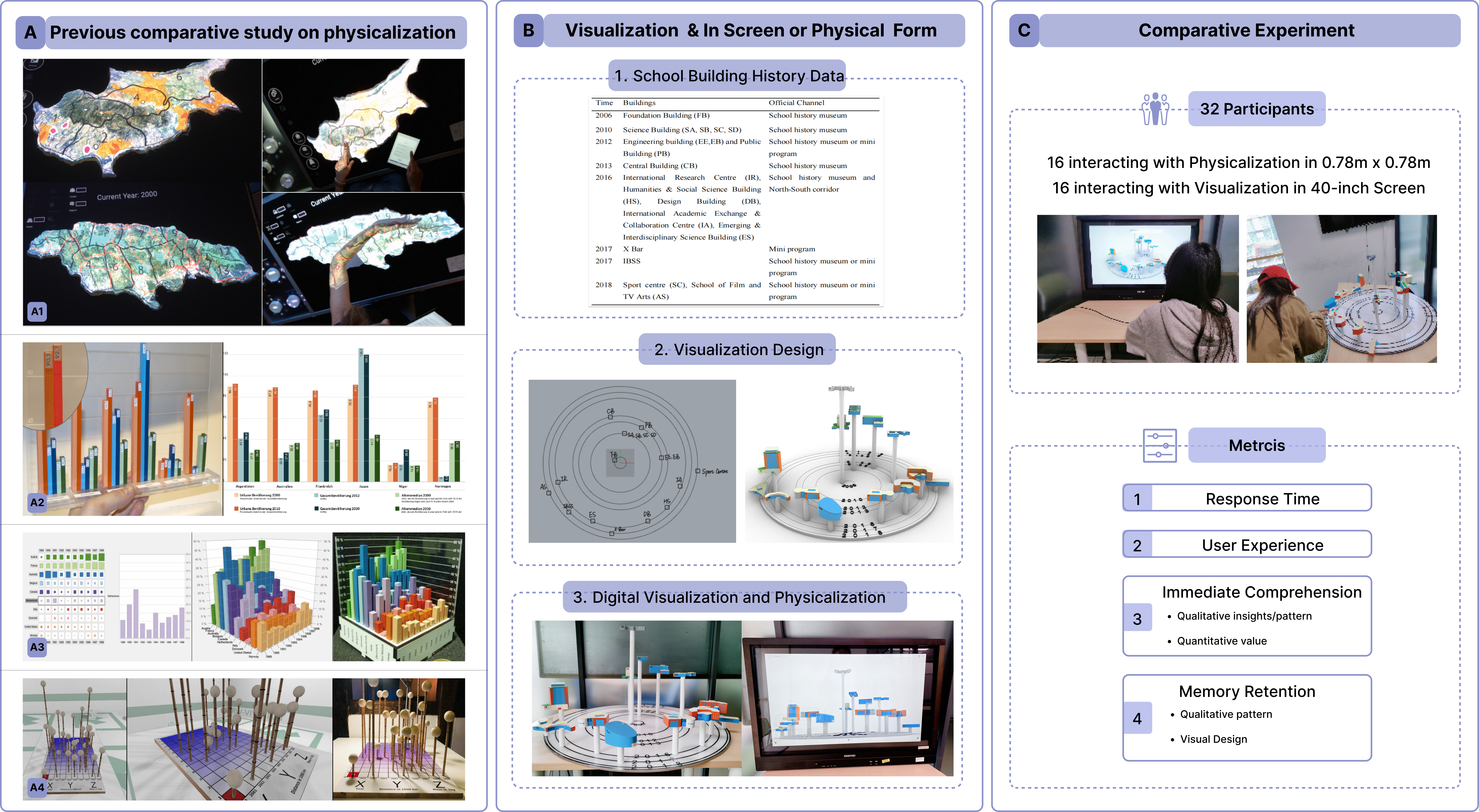} 
\caption{Overview of the experiment: \textbf{A} Experiment trigger: Previous research on physicalization (\textbf{A1} Projected 2D surface \textbf{versus} 3D terrain \cite{Kirshenbaum:2020:DCE} \textbf{A2} Physical Visualization \textbf{versus} Digital Visualization \cite{Stusak:2015:EMP} \textbf{A3} Physicalization \textbf{versus} On-screen Visualization \cite{Jansen:2013:EEP} \textbf{A4} Physicalization \textbf{versus} Virtualization \cite{Ren:2021:CUM}); \textbf{B} The process of achieving visualization; \textbf{C} Experimental design and evaluation metrics}
\label{fig:overview}
\end{figure*}

In an era characterized by an overwhelming abundance of information, the role of visualization has become pivotal in ensuring the effective comprehension and retention of data. As individuals and organizations generate and consume vast volumes of data, the transformation of this data into accessible and memorable formats is crucial for cognitive development and adaptive capabilities \cite{Manyika:2011:BIG}. Visualization, as a unique method of representing data, plays a central role in this process, enabling a deep and intuitive understanding of complex datasets \cite{Manovich:2010:V, Van:2005:VV}.

Recent technological advancements, such as 3D printing and laser cutting \cite{Ang:2019:PCB, Djavaherpour:2021:DPS}, have enabled the creation of physical models with greater precision and speed, thus enhancing the accessibility of physicalization in visualization. Physicalization, which involves the presentation of data on tangible models, has been increasingly explored and applied in visualization fields. This medium has shown potential to complement or even surpass digital forms in certain scenarios. For instance, Swedish global health specialist Hans Rosling used physical commonplace items (LEGO blocks, toilet paper rolls, etc) to illustrate data regarding global development issues in a TED talk \cite{Dragicevic:2020:DP}, which allows the audience to capture and understand these professional data in a short period of time. 
Also, physicalization was used to show four-dimensional blood flow data \cite{Ang:2019:PCB} and the findings revealed that physicalization enhanced the audience's understanding of the data in a more lucid and straightforward manner, enabling them to swiftly compare data relationships among various segments. 
 
Despite the growing prevalence of physicalization, our understanding of its effectiveness remains limited. Currently, the data physicalization research lie in its focus on small-scale and simple visual forms, such as bar charts \cite{Stusak:2015:EMP}
 and maps \cite{Kirshenbaum:2020:DCE}, typically not exceeding dimensions of $0.6$ square meters \revision{(hand-held)} \cite{Jansen:2013:EEP}. However, these findings are difficult to apply directly to larger-scale physicalizations because previous research has already indicated that size significantly affects user perception and experience of data \cite{Gramazio:2014:RVS}. Large-scale physicalizations are becoming increasingly common in various contexts, such as teaching students the timing of significant historical events \cite{Dragicevic:2020:DP}, displaying meteorite landing data in museums \cite{Ren:2021:CUM}, and presenting abstract data to help audiences understand the content \cite{Dragicevic:2020:DP}.

This study aims to bridge this gap by investigating the efficacy of large-scale data physicalizations compared to screen-based digital forms. Also, we examine complex visualizations as opposed to simpler forms like bar charts \revision{\cref{fig:overview}-A2,A3} and maps \revision{\cref{fig:overview}-A1}. Specifically, our research compared two table-sized visualization media on their impact on user experience, immediate comprehension, and memory retention using campus building history data based on 3D annual ring shapes. Through this comparative experiment, we seek to provide valuable insights for designers and developers in selecting appropriate visualization and display mediums. Ultimately, this research aims to enhance the public's ability to acquire information and learn in communal spaces (see study overview in \cref{fig:overview}).

Our contributions to the field of data visualization are multifaceted. Firstly, we introduce a novel design for data physicalization based on spatiotemporal data. This design communicates the intended information clearly and intuitively. Secondly, our research provides a robust comparative analysis of digital and physical data representations. We conducted a comprehensive evaluation on \revision{four} dimensions—\revision{response time,} user experience, \revision{information} comprehension, and memory retention—employing both qualitative and quantitative methods. This dual approach allows us to draw nuanced insights into how different mediums affect the absorption and recall of information, providing a detailed understanding of the cognitive and perceptual dynamics at play. Lastly, the reflections from our study represent our third major contribution. These recommendations offer practical guidance for future designers and developers on choosing the most effective mediums for data visualization. By outlining the strengths and limitations observed in our experiments, we equip practitioners with the knowledge to enhance the efficacy of educational and public displays, thereby improving public learning and information dissemination in various settings. These contributions collectively advance the field of data visualization, pushing forward the boundaries of how data is represented and interacted with in both digital and physical forms.

\section{Related Work}
Our work builds upon previous research on data physicalization and its comparison with other visualization methods.

\subsection{Data Physicalization}
Data physicalization transforms abstract data into tangible forms, enhancing haptic exploration and problem-solving \cite{Hull:2017:BDA}. This method is widely used across various fields. For instance, in manufacturing, General Motors head engineer Kevin Quinn uses a customized Lego-based board to monitor and update the progress of the manufacturing line, motivating engineers by making their work visible \cite{Jansen:2015:OCD}. In the medical field, physical models help patients understand their conditions better, such as using physical visualizations to describe hip pain or to analyze four-dimensional blood flow data from cardiac MRI \cite{ Ang:2019:PCB, Thudt:2018:SPP}. In education, physical visualizations enhance students' creative and structural thinking, allowing them to demonstrate lifestyle habits like sleep duration and running routes through tangible models \cite{Anderson:2017:PFD, Menheere:2021:LDD, Perin:2021:SLP}.

Physical visualizations can be static or dynamic. Static models, such as bar charts, are straightforward and simple, while dynamic models offer more interactive experiences. For example, the EMERGE bar chart includes push-pull operations and RGB data output, providing a more immersive experience through direct haptic interaction \cite{Taher:2015:EIP, Taher:2016:IUD}. Dynamic microrobots, like Zooids, represent data points that can move and be manipulated in real-time, aiding in tasks like selecting job applicants or tourist destinations \cite{Le:2018:DCD}. Effective design of physical visualizations involves using representative and symbolic materials (e.g., pencils for study time, forks for meals) \cite{Perin:2021:SLP}, adding color to enhance comprehension \cite{Gwilt:2012:EUS}, and ensuring readability and engagement while minimizing user interaction \cite{Bhargava:2017:DSP, Van:2005:VV}. Medium-sized visualizations are particularly effective for capturing data, while larger sizes leave a lasting visual impact \cite{Lopez:2021:SDP}. This comprehensive approach to design makes data physicalization a powerful tool for conveying complex information and engaging users meaningfully.

\subsection{Comparative Study on Data Physicalization}

In recent years, the performance of physicalization in practical applications has emerged as a major topic of investigation in the field of visualization research. Researchers have devoted their efforts to comparing physicalization with various visualization methods across multiple dimensions (\cref{previous literature comparison} in Appendix), aiming to comprehensively understand its performance and effectiveness in practical applications. 

\noindent\textbf{Task Completion Time.} Jansen et al. \cite{Jansen:2013:EEP} compared a 3D physical visualization with multiple on-screen models, including VR. Their findings indicated that physical visualization significantly outperformed the on-screen models in terms of task completion time. This was attributed to the direct hand interaction between participants and the physical visualization, which facilitated quicker task execution. However, the physical visualization used in this experiment was limited in size, measuring only 8 cm, which posed constraints in displaying multiple bar chart data.
Also, Ren et al. \cite{Ren:2021:CUM} compared physical visualizations with VR counterparts and found that physical visualizations significantly outperformed VR in terms of task completion time due to the tangible interactions they afford. 

\noindent\textbf{Memorability.} Stusak et al. \cite{Stusak:2015:EMP} provided insights into the comparative memorability of physical and digital visualizations using bar charts. The study found that physicalization exhibited higher memorability for long-term memory tasks, showing lower forgetting effects. However, physicalization performed relatively poorer in immediate memory tasks. Stusak suggested that the size limitation of the physical visualization (28 cm) used in the experiment could be a contributing factor. The handheld physical visualization reduced the extent of haptic engagement and interaction, potentially impacting immediate recall.

\noindent\textbf{Interactivity and Usability.} Kirshenbaum et al. \cite{Kirshenbaum:2020:DCE} extended the comparison to topographic maps, evaluating geographical information projections on 3D and 2D map models. They found that 3D visualizations demonstrated greater advantages in terms of interactivity due to the enhanced engagement afforded by the physical medium. However, in terms of usability, 2D representations were clearer and more comprehensible for users. This highlights the trade-off between the enhanced interactivity of 3D physical visualizations and the clarity provided by 2D digital visualizations. 

\subsection{Research Gap and Our Study}
While these studies provide valuable insights into the benefits and limitations of physicalization, they primarily focus on relatively small-scale models. Our research aims to fill this gap by comparing larger-scale physicalizations (such as table-sized models) with similar-sized digital displays. We investigate how these larger physical (the scale of the furniture, which ensuring stability and portability for evaluating the constructions in public settings \cite{Lopez:2021:SDP}.) and digital visualizations impact information understanding and memory retention.

Our study uses a 0.78-meter diameter physical model and a similarly sized electronic screen to explore these differences. By focusing on larger visualizations, we aim to provide more relevant insights for applications in public and educational settings, where understanding and memorability of information are crucial. This approach helps extend the current understanding of data physicalization to more practical, real-world applications involving larger-scale visualizations.

\section{Data and Visualization Design}
This section provides an overview of the data selected for visualization and the design choices made for both physical and digital representations. The aim of this study is to evaluate how different visualization mediums affect \revision{response time} user experience, information comprension and memory retention. We used publicly available data on our university buildings' completion dates due to their intrinsic interest and relevance to our participants. By utilizing both physical and digital visualization methods, we extend previous research focused on simpler forms like bar charts \cite{Jansen:2013:EEP,Stusak:2015:EMP} and maps \cite{Kirshenbaum:2020:DCE}, enabling a deeper exploration of more complex data visualizations.

\subsection{Data Collection}

We selected the completion dates of our university's main campus buildings as the visualized data. 

This data choice was guided by several key considerations:

\begin{itemize}[leftmargin=*, itemsep=0.1em]
    \item \textbf{Relevance to Participants:} Our primary user group consists of students who are readily accessible and have a vested interest in the university's history. By selecting data that they find interesting yet may not easily obtain themselves, we aimed to increase their engagement and motivation.
    \item \textbf{Educational Value:} Understanding the chronological and spatial development of campus buildings helps students and long-term faculty better connect with the university's history and evolution. Even experienced campus community members often lack detailed knowledge of these aspects, making this data both informative and engaging. By visualizing the development of the university's infrastructure, we can provide a meaningful educational experience that strengthens participants' connection to the institution.
    \revision{\item \textbf{Complexity of Data:}  Unlike previous studies that focused on either purely spatial information or simple numerical data represented in bar charts \cite{Jansen:2013:EEP}, our data combines temporal and spatial elements, making it more complex. This complexity provides a richer, more detailed understanding of the effectiveness of different visualization media.}
\end{itemize}

We collected the completion years for these buildings (\cref{collected data}) through three official sources: the school history museum, the exhibition map in the North-South corridor, and the official WeChat mini-program.

\begin{table}[t]
    \centering
    \caption{Year of Initial Building Utilization}
    \label{collected data}
    \renewcommand{\arraystretch}{1.2}
        \begin{tabular}{lp{.37\textwidth}l}
            \toprule
            \textbf{Time} & \textbf{Buildings} \\
            \midrule
            2006 & Foundation Building (FB) \\
            2010 & Science Building (SA, SB, SC, SD) \\
            2012 & Engineering building (EE,EB) and Public Building (PB) \\
            2013 & Central Building (CB) \\
            2016 & International Research Centre (IR), Humanities \& Social Science Building (HS), Design Building (DB), International Academic Exchange \& Collaboration Centre (IA), Emerging \& Interdisciplinary Science Building (ES) \\
            2017 & X Bar \\
            2017 & International Business School Suzhou (IBSS)  \\
            2018 & Sport Center (SC), School of Film and TV Arts (AS) \\       
            \bottomrule
        \end{tabular}
\end{table}

\subsection{Visualization Design}

The visualization media served as the sole independent variable in our experiment, focusing on both three-dimensional (3D) physical and digital visualizations to assess their impact on \revision{response time}, user experience, data comprehension, and memorability.

\begin{figure}[ht]
\centering
\includegraphics[width=0.45\textwidth]{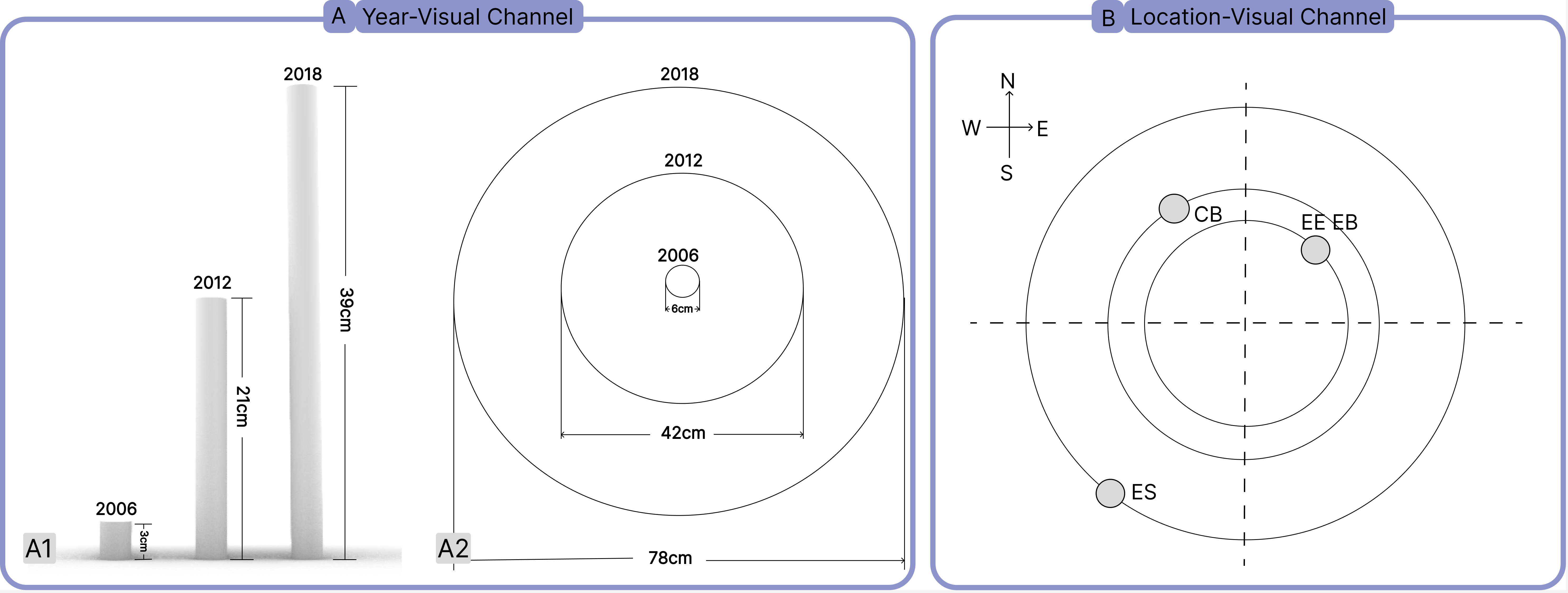} 
\caption{Encoding details in our visualization: \textbf{A} Two channels for encoding the year-visual (\textbf{A1} Height of buildings: the smallest height is $3cm$ and the highest is 39cm. It increases by $3cm$ with each additional year \textbf{A2} Diameter of the "tree rings": the smallest diameter is $6cm$ and the largest is 78cm. It increases by $6cm$ with each additional year); \textbf{B} One channel for encoding the location-visual (Position placed on the "tree rings")}
\label{fig:data encode}
\end{figure}

\subsubsection{Physicalization} 
We adopted the ``tree ring'' format to visualize the completion years of the 14 university buildings (as seen in \cref{fig:overview}-B2). Each ring represents a specific year (from 2006 to 2018), and the positions of the buildings on the rings maintain their relative spatial relationships. For instance, buildings on the same side of a ring belong to the same campus area (either the North or South Campus).
To avoid issues with displaying multiple buildings at the same height, we also encoded the historical duration of each building using the length of the cylindrical supports beneath them. Thus, from \cref{fig:data encode}-A, we can see the overall visualization encodes the building's history through the distance from the model's center and the cylinders' height. The positioning on each ring approximately preserves the original relative locations of the buildings.

Our design leverages a familiar time representation format (tree rings) and effectively uses two encoding channels to display the buildings' completion sequence and relative positions. The primary design goal is to enhance user engagement and interest, enabling a quick and intuitive understanding of the campus's architectural history while meeting the aesthetic requirements common in artistic installations.

Regarding size, we chose a \revision{large} scale (diameter of 0.78 meters, height of 0.39 meters) commonly seen in museum and exhibition settings. This table-sized format balances the advantages of both large and small-scale visualizations, allowing users to view the entire visualization and inspect details easily \cite{Lopez:2021:SDP}. The physical model consists of a circular base with cylindrical supports and miniature building models on top. The buildings' colors closely matched their real-life counterparts on campus, enhancing the model's authenticity and visual appeal \cite{Gwilt:2012:EUS}.

\subsubsection{Digitalization} 
For digital visualization, we used Rhino software to ensure consistency with the physical one's colors and design \cite{Gwilt:2012:EUS}. The Rhino-rendered digital model provides a 3D multi-perspective view (as shown in \cref{fig:teaser}-A), maintaining the same spatial relationships and color schemes as the physical model.

To ensure an equitable comparison, we displayed the digital visualization on a touch-enabled electronic screen of similar size (40-inch; $0.88m\times0.49m$) to the physical model. We also keep the display view the same as the physical visualization size. This setup included two common interaction methods in museums and art exhibitions: direct touch interaction on the screen and interaction via a smaller connected touchscreen device. For this study, we connected a 10.9-inch tablet to allow users to experience different interaction modes and choose their preferred method. Participants were allowed to observe from any angle and height for physicalization, ensuring consistent viewing experiences.

\section{Experiment}
\subsection{Participant and Experimental Design}

\begin{figure*}[htb]
\centering
\includegraphics[width=0.95\textwidth]{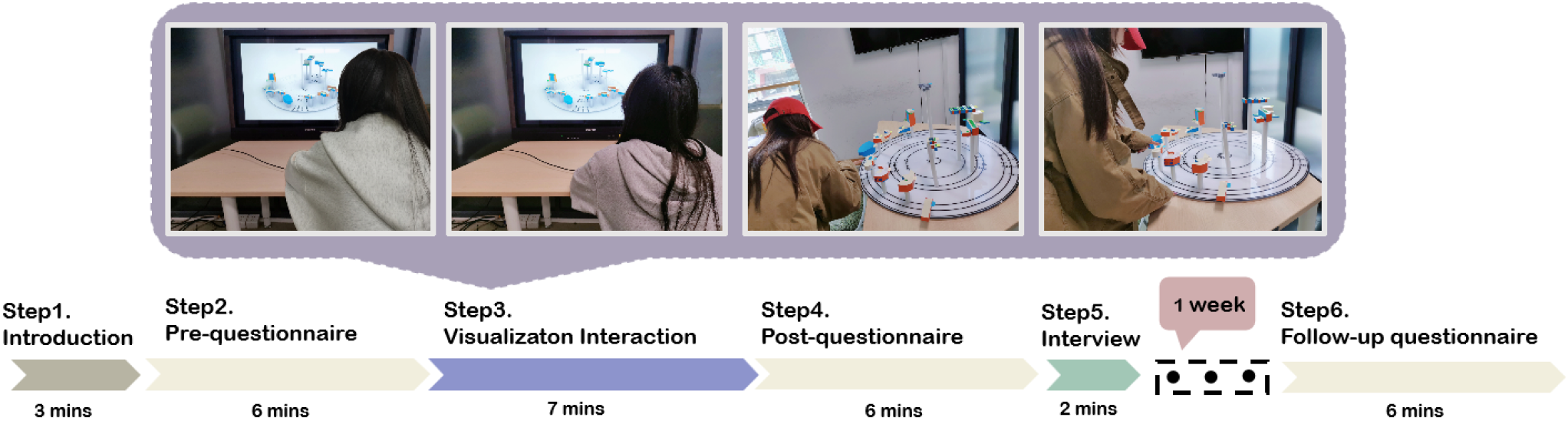} 
\caption{Overview of the entire experimental procedure}
\label{fig:experience process}
\end{figure*}

\revision{In our study, we had 32 participants, with each participant assigned to one of the two conditions in a between-subjects design. Using G-power \cite{Serdar:2021:SSP} for an independent two-sample t-test with an effect size of 0.8, an alpha level of 0.05, and a power of 0.85, the required sample size was approximately 29 participants. Our sample size of 32 participants (16 per condition) is adequate and slightly exceeds the required number, ensuring sufficient power to detect significant effects.
They were reached through multiple social media }at our university, ranging in age from 19 to 26 years, with an average age of 22. The group included 1 doctoral student, 1 graduate student, and 30 undergraduate students. These participants are mainly from five different backgrounds, including computer science (18), business (10), engineering (2), mathematics (1), and environment (1).

The participants were divided into two equal groups, each consisting of 8 males and 8 females. \revision{ Both groups of participants were conducted within the same laboratory setting, where the individuals had unrestricted access to engage with models in the designated area, allowing for free interaction during the experiment.}

\noindent\textbf{Group A:} This group observed the digital visualization displayed on a 40-inch LCD touch screen (\cref{fig:teaser}-B1). To control for the variable of interest—the type of visualization—participants were only allowed to rotate the model but not zoom in or out directly or using an iPad. This ensured that any differences in memorability could be attributed to the visualization method rather than variations in model size.

\noindent\textbf{Group B:} Participants in this group were tasked with observing the physical visualization (\cref{fig:teaser}-B2). They were given complete freedom to interact with the model, including touching it and moving around to view it from different angles. This unrestricted interaction was intended to provide a comprehensive understanding of the physical model's impact on memorability.

\subsection{Metrics}

The specific metrics of our experiment included response time, user experience,  immediate feedback, and long-term memory retention measured seven days later.

\noindent\revision{\textbf{Response Time.} To effectively evaluate the impact of two different media on participants' memorability regarding questions of varying difficulty levels, the duration spent contemplating each participant's responses to different levels of questions was documented.  A timer was used to track how long participants spent on each question.
}

\noindent\textbf{Quantitative Data Comprehension and Recall.} 
The quantitative assessment of data comprehension and recall was divided into three task-based categories inspired by map-related tasks \cite{Roth:2013:ETI, Shneiderman:2011:CSC}, as our data combines temporal and spatial elements. These tasks were designed to measure different levels of cognitive engagement \revision{(see details in \cref{QA Experimental Questionnaire})}:
\begin{enumerate}[leftmargin=*, itemsep=0.03em]
    \item [(1)]\textbf{Identification (Easy)}: Participants were asked to identify and recognize individual buildings based on their location and completion year.
    \item [(2)] \textbf{Comparison (Moderate)}: Participants were required to compare two buildings, determining which one was completed first or relative position.
    \item [(3)] \textbf{Ranking (Difficult)}: Participants were tasked with ranking three buildings according to their completion dates or ranking their relative position from north to south or from east to west.
\end{enumerate}

The assessment questionnaires for both the year and geographical location data included 12 questions each, with four questions per difficulty level. This structure allowed us to evaluate accuracy and response time across different levels of complexity.

\noindent\textbf{Qualitative Pattern Comprehension and Recall.}
In addition to quantitative tasks, participants were asked to provide qualitative feedback on the patterns or trends they noticed in the visualizations. We recorded these observations immediately after the visualization session and revisited them during the long-term memory test seven days later. Participants were also required to describe or sketch their memory of the visualization, helping to capture the depth and persistence of their recall.

\noindent\textbf{User Experience.} 
The user experience was assessed using the User Experience Questionnaire (UEQ) \cite{Rauschenberger:2013:EMU, Schrepp:2014:AUE}, which is normally used to evaluate interactive products based on various criteria such as attractiveness, perspicuity (\revision{shown in} \cref{user Experiment question}), efficiency, dependability, stimulation, and novelty. Participants completed this questionnaire immediately after interacting with the visualization, capturing their immediate reactions and experiences.

\subsection{Experiment Process}
To comprehensively understand how visualization methods affect immediate comprehension and long-term memory retention, we divided the evaluation into three key stages: pre-test, immediate, and long-term (a week later). This design allowed us to assess participants' instant memory of the data and observe memory decay over time, aligning closely with numerous previous studies on the memorability of visualizations \cite{Chettaoui:2023:UMT, Saket:2016:BUP, Shneiderman:2011:CSC, Stusak:2015:EMP}. A 7-day interval for long-term testing is considered appropriate and has been validated in prior research as an effective timeframe for assessing long-term memory retention \cite{He:2024:VHT, Zdanovic:2022:IDS}. Our study was approved by the school's ethical board before its initiation.

\begin{enumerate}[leftmargin=*, itemsep=0.1em]
    \item [(1)] \textbf{Introduction and Consent:} 
    \begin{itemize}[leftmargin=*, itemsep=0.1em]
        \item \textbf{Project Introduction:} The experiment began with an introduction of the project, explaining the study's objectives, methodology, and significance.
        \item \textbf{Informed Consent:} Participants were provided with an information sheet and consent form to ensure they understood the study and their rights.
        \item \textbf{Pre-Test Questionnaire:} Participants completed a background information and knowledge questionnaire, similar to \cite{Chettaoui:2023:UMT, He:2024:VHT}. This included questions about their background information and prior knowledge of the subject matter to ensure a baseline understanding before the experiment.
    \end{itemize}
    
    \item [(2)] \textbf{Observation and Interaction:}
Participants were assigned to either the digital or physical visualization group, with 7 minutes to explore their assigned visualization. This duration was determined through multiple rounds of testing among the authors, ensuring it was an optimal time for most participants to engage with and remember the information effectively.
        \begin{itemize}[leftmargin=*, itemsep=0.1em]
            \item \textbf{Digital Group:} Interacted with the digital model on a large touch screen or small iPad, rotating it to view different angles.
            \item \textbf{Physical Group:} Touched and viewed the physical model from various perspectives, moving around it freely.
        \end{itemize}

    \item [(3)] \textbf{Immediate Testing:}
Participants completed a questionnaire to assess their understanding and memory after a 7-minute interaction.
        \begin{itemize}[leftmargin=*, itemsep=0.01em]
            \item \textbf{Questionnaire Structure:} The survey was divided into six sub-sections, covering year and location data across three difficulty levels.
            \item \textbf{Insight and Pattern Recognition:} Participants wrote down approximately three insights or patterns they observed.
            \item \textbf{User Experience Questionnaire:} Participants also completed a User Experience Questionnaire (UEQ).
        \end{itemize}

    \item [(4)] \textbf{Long-term Memory Testing:}
  One week later, participants took a similar test to assess long-term memory retention.
        \begin{itemize}[leftmargin=*,itemsep=0.05em]
            \item \textbf{Questionnaire Consistency:} The questionnaire content was identical, with the question order shuffled.
            \item \textbf{Visualization Recall:} Participants were asked to draw or write down their recollection of the visualization.
            \item \textbf{Offline/Online Options:} Participants could complete the test offline or online.
        \end{itemize}

\end{enumerate}
To examine significant differences between the data from different models, we employed the t-test method. Additionally, to investigate memory retention over different time points within the same group, we conducted repeated-measure ANOVA. This approach allowed us to assess whether memory retention significantly impacted the data across the three testing sessions.

\section{Results}

Our comparative study examined the effectiveness of digital and physical visualizations across three key metrics: response time, user experience, and immediate comprehension and long-term memory retention of information.  Below, we summarize and analyze our findings in these areas.

\subsection{Response Time}

We measured and recorded participants' time to complete quantitative information comprehension questions. 

A t-test revealed no statistically significant differences in response times between the two visualizations, with p-values of 0.605.
Participants who observed the digital visualization (Group A) had an average response time of 149.91 seconds ($SD=73.58$), and participants who observed the physical visualization (Group B) had an average response time of 143.06 seconds ($SD=70.41$).

\subsection{User \revision{E}xperience}

\begin{figure}[h]
    \centering
    \includegraphics[width=0.40\textwidth]{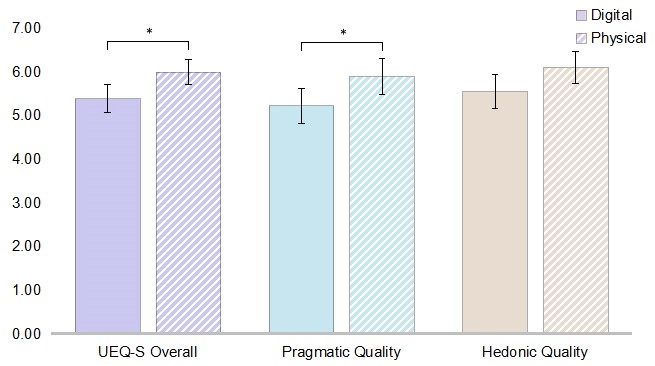} 
    \caption{Results of the mean score of the two visualizations on user experience. (* indicates a significant difference)}
    \label{fig:user experience result}
\end{figure}

We conducted an analysis of the user experience with a confidence interval set at 95\%. The findings (see \cref{fig:user experience result}) indicated a significant difference between the two types overall (p = 0.011), particularly in terms of pragmatic quality (p = 0.030). However, no significance was observed in hedonic quality.
These results suggest that while both visualizations were similarly enjoyable, participants found the physical visualization more practically useful.

\subsection{Information \revision{C}omprehension}

\noindent \textbf{Quantitative Results:}
Before analyzing accuracy, participants' questionnaire responses were encoded: correct answers were coded as 1, while incorrect answers and ``I don't know'' responses were coded as a 0. 95\% confidence interval was applied. 

We assessed participants' comprehension of information three times: before using the visualizations, immediately after use, and seven days later. The same set of questions was used for each test, with the order randomized. The \revision{repeated-measure ANOVA} revealed significant differences in accuracy rates across the three different times for each medium (p = 0.001 for physicalization, p = 0.007 for digitalization).
For the digital visualization group, the average pre-test accuracy was 31.25\% ($SD=0.148$), which increased to 78.13\% ($SD=0.163$) immediately after use and retained a rate of 58.85\% ($SD=0.141$) after seven days. For the physical visualization group, the average pre-test accuracy was 35.16\% ($SD=0.149$), which improved to 84.90\% ($SD=0.115$) immediately after use, and retained a rate of 73.96\% ($SD=0.157$) after seven days. The detailed results can be seen in \cref{total accuracy}.
We conducted t-tests to compare the accuracy rates between the two visualizations at each time point. The pre-test accuracy differences were not statistically significant (p = 0.089), indicating that both groups started with similar levels of knowledge. However, significant differences were found in the accuracy rates immediately after using the visualizations (p = 0.042) and seven days later (p = 0.028), with the physical visualization group consistently outperforming the digital group.

\begin{figure*}[ht]
    \centering
    \includegraphics[width=0.98\textwidth]{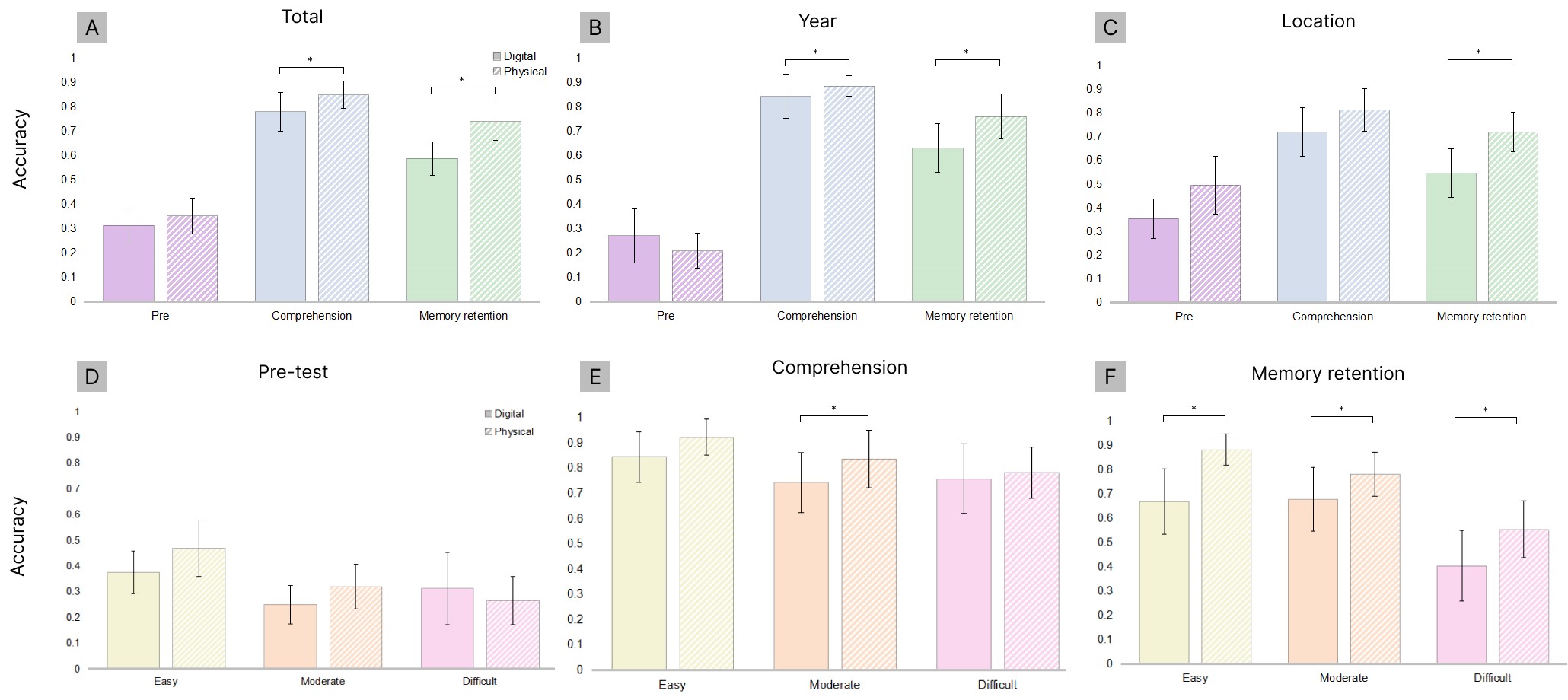} 
    \caption{Average accuracy of the two visualizations at different test time and problem levels: \textbf{A} Total average accuracy; \textbf{B} Average accuracy of the year; \textbf{C} Average accuracy of the location; \textbf{D} Average accuracy of three-level questions in pre-test; \textbf{E} Average accuracy of three-level questions in comprehension; \textbf{F} Average accuracy of three-level questions in memory retention. (* indicates a significant difference)}
    \label{fig:accuracy results}
\end{figure*}

Further analysis was conducted to examine the accuracy for different types of data (year and location) and across different difficulty levels (\cref{fig:accuracy results}):

\begin{itemize}[leftmargin=*]
    \item \textbf{Year:} The average pre-test accuracy was 27.08\% ($SD=0.227$) for the digital visualization and 20.83\% ($SD=0.149$) for the physical visualization, showing no significant difference initially. After using the visualizations, the immediate memory accuracy increased to 84.38\% ($SD=0.185$) for the digital group and 88.54\% ($SD=0.085$) for the physical group, with a significant difference between the two (p = 0.023). Seven days later, the retention accuracy was 63.02\% ($SD=0.204$) for the digital group and 76.04\% ($SD=0.187$) for the physical group, with a significant difference (p = 0.025).
    \item \textbf{Location\revision{:}} The average pre-test accuracy was 35.42\% ($SD=0.171$) for the digital visualization and 49.48\% ($SD=0.252$) for the physical visualization. Immediate memory accuracy improved to 71.88\% ($SD=0.211$) for the digital group and 81.25\% ($SD=0.181$) for the physical group. Seven days later, the retention accuracy was 54.69\% ($SD=0.209$) for the digital group and 71.88\% ($SD=0.172$) for the physical group, with a significant difference (p = 0.041).
    \item \textbf{Difficulty Levels:} Analysis of specific recall questions at different levels showed significant differences in the moderate questions (p = 0.026). Seven days later, the retention accuracy showed significant differences across easy, moderate and difficult questions (p = 0.043, p = 0.032 and p = 0.038, respectively). 
\end{itemize}

\noindent \textbf{Qualitative Results:}
Participants were asked to summarize the patterns they observed from the visualizations immediately after usage (see details in \cref{pattern_digital} and \cref{pattern_physical} in Appendix). 
Among the digital visualization group, only 5 participants mentioned relevant patterns, while \revision{8} participants from the physical visualization group identified these patterns. The main patterns observed included:
(1) a closer proximity between buildings correlates with a shorter duration until they are put into use, (2) the North campus was initially operationalized, followed by the subsequent utilization of the South campus, and (3) the buildings in the South campus were occupied at a faster pace compared to the North campus.

\begin{figure*}[htb]
    \centering
    \includegraphics[width=0.95\textwidth]{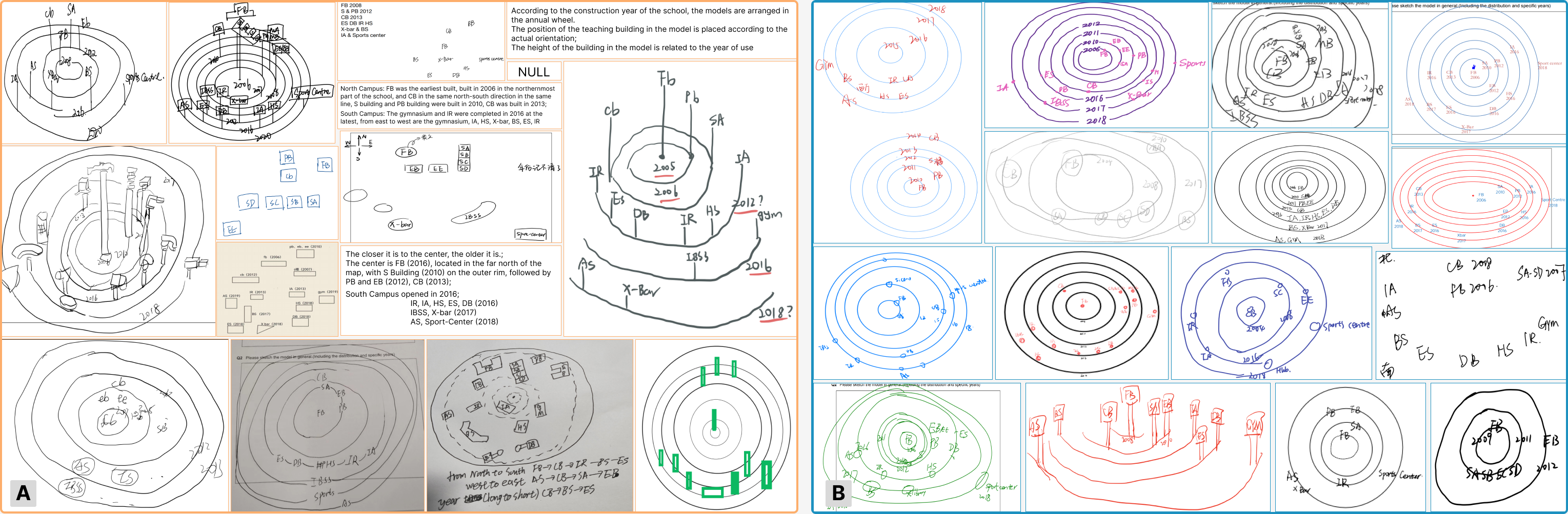} 
    \caption{The sketch drawn by participants: \textbf{A} Digital Visualization (Group A); \textbf{B} Physical Visualization (Group B).}
    \label{fig:sketch}
\end{figure*}

Seven days later, participants were again asked to recall the patterns. Consistently, participants who used the physical visualization could recall more patterns accurately. Additionally, they were asked to accurately sketch the visualization they had used (see \cref{fig:sketch}). In the digital group, only \revision{8} participants could provide a rough depiction of the tree-ring visualization and spatial distribution, whereas \revision{14} participants in the physical group successfully completed the drawing task satisfactorily.

\subsection{Summary}

Our results indicate that while there are no significant differences in response times between digital and physical visualizations, physical visualizations offer a better user experience and significantly improve immediate comprehension and long-term memory retention of information. The main differences between the two visualizations are evident in more difficult tasks, where physical visualizations show a clear advantage. Additionally, in terms of visual recall, physical visualizations are more effective at helping users remember the visualization itself. Furthermore, in qualitative feedback, users of physical visualizations are more likely to uncover underlying patterns within the whole dataset.

\section{Discussion}
In this section, combining the previous literature findings related to materialization, we analyze, discuss and reflect the data results.

\subsection{Discussion on Results}
\begin{itemize}[leftmargin=*, itemsep=0.2em]
\item \textbf{Response Time\revision{.}} Our study found no significant difference in response time between the two visualization media. This finding contrasts with Jansen's study, which reported faster task completion times for \revision{small-sized ($8cm\times8cm$)} physical visualizations compared to on-screen models \cite{Jansen:2013:EEP}. \revision{When the medium size is large, there is no significant difference in effectiveness between the screen and physical forms.}
\item \textbf{User Experience\revision{.}} The user experience analysis showed that physicalization outperformed digital visualization. This aligns with findings from previous studies, such as Stusak et al. \cite{Stusak:2015:EMP}, which found that physical visualization demonstrated superiority over the tablet to provide a more enjoyable reading experience for participants when presenting bar chart-related data. 

\item \textbf{Immediate Comprehension\revision{.}}
Our analysis revealed a significant effect of display media on immediate comprehension, with physical visualizations demonstrating higher accuracy compared to digital ones. We attribute this outcome to two main factors. First, physical visualizations enable participants to directly interact with the data, engaging in a more immersive exploration through tactile means and examining the data from multiple perspectives. This direct interaction likely enhances the capture and memorization of knowledge more effectively \cite{Djavaherpour:2021:DPS, Hull:2017:BDA}. Additionally, the unique form and appearance of our 3D visualizations are more easily perceived in physical form, facilitating immediate comprehension. The tangible nature of physical models allows participants to relate to and recall the information more easily, as suggested by Dragicevic \cite{Dragicevic:2020:DP}. These factors combined suggest that physical visualizations offer a superior means for users to quickly and effectively comprehend information compared to screen-based displays.

\item \textbf{Memory Retention\revision{.}}
In the assessment of data retention one week later, the physicalization exhibited superior accuracy compared to digital ones. 
Nearly all participants who observed the physicalization could accurately depict the approximate design outline (tree rings), supporting López's conclusion that physical visualizations leave a more profound impression on the audience \cite{Lopez:2021:SDP}.
This finding also corroborates Stusak's research \cite{Stusak:2015:EMP} on memory retention for physical and on-screen bar chart data, which indicated that physicalization tends to mitigate memory loss over time. 
The tactile and visual engagement with physical models can reinforce memory encoding, leading to better long-term retention of complex data.  Also, this enhanced retention can be attributed to physical models' clear spatial information and interconnectivity, which help individuals establish stronger spatial relationships and associations in their memory \cite{Hull:2017:BDA}. 
\end{itemize}

\subsection{Reflections}
Here are some key reflections and insights drawn from our work:
\begin{itemize}[leftmargin=*, itemsep=0.2em]
\item \textbf{Comprehensive Data Collection and Analysis.} 
One notable aspect of our experiment was the extensive data collection, encompassing quantitative and qualitative measures. We evaluated users' comprehension through specific data comparisons, sorting tasks, identification, and the ability to discern patterns and underlying insights. This approach contrasts with many prior comparative studies that often focus solely on basic quantitative metrics.
Our study was influenced by research \cite{Niklas:2012P:patterns} suggesting that the primary purpose of visualization is not merely to recall specific values but to understand and interpret broader data patterns and trends. As noted, effective visualization facilitates insight generation, allowing users to grasp relationships and patterns within the data. Therefore, including pattern recognition tasks in the evaluation was a critical design choice to capture a more holistic understanding of visualization effectiveness.

However, it is important to acknowledge that our pattern recognition analysis remains at an early stage, primarily counting the frequency of identified patterns. Additionally, because many participants lacked formal training in visualization, their reflections might have focused more on the visual encoding rules rather than the underlying data patterns (see \cref{pattern_digital} and \cref{pattern_physical} in Appendix). Future studies should develop more sophisticated methods for analyzing and summarizing these insights and explore ways to guide users better or employ alternative methods to elicit meaningful patterns and insights from visualizations.

\item \textbf{Interaction Differences and Their Impact.}
Another significant reflection involves the interaction differences between the two visualization media. Physicalization allows for a more immersive and active exploration, with users physically moving around and viewing the model from various angles. This tactile interaction often results in a more engaging and memorable experience. \revision{Inversely}, digital visualizations rely on screen-based interactions, such as rotating the model on a touch screen. Although we attempted to mitigate interaction complexity by providing an iPad for more manageable navigation, users still preferred the direct, hands-on interaction offered by physicalization.

This preference highlights an important consideration for selecting visualization media: the complexity and nature of user interaction. Physicalizations offer more intuitive and satisfying interactions, which can significantly enhance user experience. However, when interaction complexity is minimized, as in our study, the differences in user experience between the two media may be reduced. Despite our efforts to simplify digital interactions, physicalizations still provided a superior user experience.

\end{itemize}

\subsection{Limitation and Future Work}

This study has provided insightful contributions to understanding the effects of data physicalization. However, several limitations must be recognized, opening opportunities for further exploration.

\begin{itemize}[leftmargin=*, itemsep=0.1em]
    \item \textbf{Demographic Specificity.}
Our study predominantly involved participants with a relatively young average age. This demographic focus limits the generalizability of our findings to young individuals who may have particular aptitudes or preferences for engaging with physicalized data. To develop a more comprehensive understanding of how different age groups and demographic backgrounds respond to data physicalization, future studies should encompass a broader range of participants. This expansion would help ascertain whether the observed benefits of physicalization hold across diverse populations and how different demographic characteristics influence the effectiveness of visualization techniques.
\item \textbf{Scale of Physical \revision{V}isualization.}
Although our study tested larger-sized physical entities and digital forms, we also observed larger-scale data physicalizations in practice, akin to the size of an entire wall or half a floor. The comparison between these grand-scale physicalizations and similar-sized digital displays remains unexplored. Investigating how the size and scale of physicalization affect user interaction and memory could provide crucial information for designing data visualizations in large public or educational spaces.
\item \textbf{Data Type and Representation.}
This research predominantly focused on static data types, specifically those related to years and geographic locations, utilizing tangible geometric models or solid materials for representation. According to visualization theory \cite{Huron:2014:CV}, the choice of medium for displaying data is often dictated by the target audience and the specific tasks at hand. In public settings, where the general audience interacts with visualizations, the complexity of these visualizations is typically kept to a minimum. Building on previous research that primarily explored simple visual forms like bar charts and conventional maps, this study ventured into a less familiar territory by experimenting with innovative spatiotemporal data visualizations.

This study initiates an important dialogue about the adaptability of data visualization strategies to accommodate more complex or varied data types. Future research could leverage this foundation to comprehensively explore the potential complexities in data visualization, aiming to enhance understanding and interaction across a broader spectrum of public and specialized audiences. Such investigations would not only enrich the field of data visualization but also refine the practical applications of these technologies in diverse real-world scenarios.

\item \textbf{Diverse Physicalization Techniques.}
Our experiment focused on most traditional methods of physicalization, yet innovative approaches, such as dynamic wheeled micro-robots \cite{Le:2018:DCD}, are emerging. Future studies could examine the effectiveness of these novel physicalization techniques compared to counterparts on the digital screen. Such research would expand the toolkit for data visualization, potentially leading to more engaging and effective ways to represent and interact with data.

\item\textbf{Longitudinal Impact.}
This study did not address the long-term effects (more than seven days) of data physicalization on memory retention. Understanding how physical and digital data representations influence memory over extended periods could provide deeper insights into their efficacy as educational and communicative tools. 
\end{itemize}

Addressing these limitations and incorporating these suggestions into future work will refine the understanding of data physicalization's benefits and broaden the scope of its applicability in real-world scenarios.

\section{Conclusion}
In this study, we compared two visualization media forms — screen-based digitization and physicalization on \revision{four} dimensions: \revision{response time}, user experience, immediate comprehension and memory retention. The experimental findings revealed a significantly positive impact of data physicalization on immediate comprehension, memory persistence and usability. 
Furthermore, distinct from prior studies which primarily focused on smaller-scale visualizations, our research emphasized the visualization on larger scales. Unlike the common bar charts or map-based data physicalizations, our study implemented more intricate visualization apparatuses. These findings have important implications for the design and application of data visualization, offering valuable guidance for the selection of table-sized display mediums in settings such as museums, art galleries, or educational tools.

\acknowledgments{
This work was supported in part by
a grant from RDF-22-01-092.}

\bibliographystyle{abbrv-doi}

\bibliography{references}

\clearpage
\appendix
\label{sec:supplemental_materials}
\onecolumn

{\LARGE\textbf{Appendix}}

\vspace{5px}
This section provides a summary of related literature, \revision{detailed questionnaires and} qualitative data from our experiment.

\section{Comparative Study on Physicalization}

\begin{table*}[htbp]
    \caption{Previous literature on physicalization comparison}
    \label{previous literature comparison}
    \renewcommand{\arraystretch}{1.2}
        \begin{tabular}[t]{p{.15\textwidth}p{.18\textwidth}p{.15\textwidth}p{.17\textwidth}p{.2\textwidth}}
        \toprule
              \multirow{2}{*}{\textbf{Author}} & \textbf{Comparison} & \textbf{Size of} & \textbf{Testing} &  \multirow{2}{*}{\textbf{Result}}\\
              ~ &  \textbf{Content} & \textbf{Physical Model} & \textbf{Metric} & ~ \\
        \midrule
            \multirow{2}{.15\textwidth}{Jansen et al (2013) \cite{Jansen:2013:EEP}} & \multirow{2}{.18\textwidth}{Physicalization \textbf{versus} Stereo \textbf{versus} Mono \textbf{versus} 2D} & \multirow{2}{.15\textwidth}{$8cm\times8cm$} & (a) Error rate; & \multirow{2}{.2\textwidth}{Physicalization performs best at time on task.} \\
             ~ &  ~ & ~ & (b) Time on task. & ~\\

             &&&&\\
             
            \multirow{2}{.15\textwidth}{Stusak et al (2015) \cite{Stusak:2015:EMP}} & \multirow{2}{.18\textwidth}{Physical Visualizations \textbf{versus} Digital Visualizations}&\multirow{2}{.15\textwidth}{$28cm\times17cm$} & (a) Correctness rate; & \multirow{2}{.2\textwidth}{Less memory loss in physicalization.} \\
             ~ &  ~ & ~ & (b) Memorability score. & ~\\

             &&&&\\

            \multirow{2}{.15\textwidth}{Ren, He \& Hornecker, Eva (2021) \cite{Ren:2021:CUM}} & \multirow{2}{.18\textwidth}{Physicalization \textbf{versus} Virtualization} &  \multirow{2}{.15\textwidth}{$60cm\times40cm\times50cm$} & (a) Response time; & (a)Lower response times in physicalization; \\
            ~ &  ~ & ~ & (b) Correctness rate. & (b) Differences of correctness rates are marginal. \\

             &&&&\\
             
             \multirow{3}{.15\textwidth}{Chettaoui et al (2023) \cite{Chettaoui:2023:UMT}} & \multirow{3}{.18\textwidth}{Tangible user interface (TUI) \textbf{versus} Classic method with non-augmented objects} & \multirow{3}{.15\textwidth}{$52cm\times70cm$} & (a) System Usability & \multirow{3}{.2\textwidth}{TUI performs better at short-term retention} \\
             ~ &  ~ & ~ & Scale (SUS) score; & ~\\
             ~ &  ~ & ~ & (b) Completion Time. & ~\\
             &&&&\\
        \bottomrule
        \end{tabular}
\end{table*}

\section{Detailed Experimental Questionnaires}

\begin{table*}[]
    \centering
    \caption{Detailed Questions and Answers of the Public Data}
    \label{QA Experimental Questionnaire}
    \renewcommand{\arraystretch}{1.3}
        \begin{tabular}{lllll}
        
        \textbf{Q1} & \multicolumn{4}{l}{Which building was/were the last to be put into use in XJTLU?} \\
        {} & \Square{} IBSS \hspace{1cm} & \CheckedBox {} Sport Centre & \Square {} DB  \hspace{1cm} & \Square {} I don’t know \\
        \textbf{Q2} & \multicolumn{4}{l}{Which building was/were the last to be put into use in the North campus of XJTLU?} \\
        {} & \Square{} SA & \CheckedBox {} CB & \Square {} EB  & \Square {} I don’t know \\
        \textbf{Q3} & \multicolumn{4}{l}{Which building was/were the first to be put into use in XJTLU?} \\
        {} & \Square{} PB  & \CheckedBox {} FB & \Square {} SD  & \Square {} I don’t know \\
        \textbf{Q4} & \multicolumn{4}{l}{Which building was/were the first to be put into use in the South campus of XJTLU?} \\
        {} & \Square{} X-bar & \CheckedBox {} ES & \Square {} AS  & \Square {} I don’t know \\

        \textbf{Q5} & \multicolumn{4}{l}{Which was used earlier, IA or AS?} \\
        {} & \Square{} AS & \CheckedBox {} IA  & \Square {} I don’t know &   \\
        \textbf{Q6} & \multicolumn{4}{l}{PB was put into use later than IBSS.} \\
        {} & \Square{} Yes & \CheckedBox {} No & \Square {} I don’t know &  \\
        \textbf{Q7} & \multicolumn{4}{l}{The year difference between SA and FB is smaller than that between CB and PB. } \\
        {} & \Square{} Yes & \CheckedBox {} No & \Square {} I don’t know &  \\
        \textbf{Q8} & \multicolumn{4}{l}{HS and IR were put into use at the same year.} \\
        {} & \CheckedBox {} Yes & \Square No & \Square {} I don’t know &  \\

        \textbf{Q9} & \multicolumn{4}{l}{Both AS and IA were put into use in 2016.} \\
        {} & \Square{} Yes & \CheckedBox {} No & \Square {} I don’t know &  \\
        \textbf{Q10} & \multicolumn{4}{l}{List the following buildings in descending order of when they were put into use. (The earliest-used building is ranked the first)} \\
        {} & \Square{} ES  & \Square {} IBSS & \Square {} Sport Centre  & \\
        {} & \multicolumn{4}{l}{$ES>IBSS>Sport Centre$} \\
        \textbf{Q11} & \multicolumn{4}{l}{Was PB put into use in 2008? } \\
        {} & \Square{} Yes & \CheckedBox {} No & \Square {} I don’t know &  \\
        \textbf{Q12} & \multicolumn{4}{l}{List the following buildings in descending order of when they were put into use. (The earliest-used building is ranked the first)} \\
        {} & \Square{} EE   & \Square {} AS & \Square {} IA  & \\
        {} & \multicolumn{4}{l}{$EE>IA>AS$} \\

        \textbf{Q13} & \multicolumn{4}{l}{ES is located in the North Campus.} \\
        {} & \Square{} Yes & \CheckedBox{} No & \Square{} I don’t know &  \\
        \textbf{Q14} & \multicolumn{4}{l}{Which is the easternmost building in the North Campus?} \\
        {} & \Square{} PB & \Square{} SC & \CheckedBox{} EB & \Square {} I don’t know \\
        \textbf{Q15} & \multicolumn{4}{l}{DB is located in the South Campus.} \\
        {} & \CheckedBox{} Yes & \Square No & \Square{} I don’t know &  \\
        \textbf{Q16} & \multicolumn{4}{l}{Which is the easternmost building in the South Campus?} \\
        {} & \CheckedBox{} Sport Centre & \Square {} HS & \Square {} X-bar  & \Square {} I don’t know \\

        \textbf{Q17} & \multicolumn{4}{l}{SD is west of PB.} \\
        {} & \CheckedBox{} Yes & \Square No & \Square{} I don’t know &  \\
        \textbf{Q18} & \multicolumn{4}{l}{In the east-west direction, which is closer to CB, FB or SC?} \\
        {} & \Square{} SC & \CheckedBox {} FB  & \Square {} I don’t know &   \\
        \textbf{Q19} & \multicolumn{4}{l}{ES is east of DB.} \\
        {} & \Square{} Yes & \CheckedBox {} No & \Square {} I don’t know &  \\
        \textbf{Q20} & \multicolumn{4}{l}{In the east-west direction, which is closer to IBSS, X-bar or Sport Centre?} \\
        {} & \Square{} Sport Centre & \CheckedBox {} X-bar  & \Square {} I don’t know &  \\

        \textbf{Q21} & \multicolumn{4}{l}{Arrange the following buildings in order from north to south.} \\
        {} & \Square{} PB \hspace{1.5cm} & \Square {} IBSS \hspace{.5cm} & \Square {} IR &  \\
        {} & \multicolumn{4}{l}{$PB>IR>IBSS$} \\
        \textbf{Q22} & \multicolumn{4}{l}{Arrange the following buildings in order from west to east.} \\
        {} & \Square{} AS  & \Square {} Sport Centre & \Square {} SB  & \\
        {} & \multicolumn{4}{l}{$AS>SB>Sport Centre$} \\
         \textbf{Q23} & \multicolumn{4}{l}{Arrange the following buildings in order from north to south.} \\
        {} & \Square{} HS & \Square {} DB   & \Square {}  SC &  \\
        {} & \multicolumn{4}{l}{$SC>HS>DB$} \\
        \textbf{Q24} & \multicolumn{4}{l}{Arrange the following buildings in order from west to east.} \\
        {} & \Square{} X-bar   & \Square {} ES & \Square {} IA  & \\
        {} & \multicolumn{4}{l}{$ES>X-bar>IA$} \\
        
        \end{tabular}
        
\end{table*}

\clearpage

\begin{table*}[]
    \renewcommand{\arraystretch}{1.3}
    \caption{User Experience about the Memorability of Data Physicalization}
    \label{user Experiment question}
    For the assessment of the product, please fill out the following questionnaire. The questionnaire consists of pairs of contrasting attributes that may apply to the product. The circles between the attributes represent gradations between the opposites. You can express your agreement with the attributes by ticking the circle that most closely reflects your impression. \\
    \\
    Example:\\
        \begin{tabular}{l|ccccccc|l}
        \hline
        attractive & \Square & \Square & \CheckedBox & \Square & \Square & \Square & \Square & unattractive\\
        \hline
        \end{tabular}
   \\
   \\
    This response would mean that you rate the application as more attractive than unattractive. \\
    \\
    Please decide spontaneously. Don’t think too long about your decision to make sure that you convey your original impression. \\
    \\
    Sometimes you may not be completely sure about your agreement with a particular attribute or you may find that the attribute does not apply completely to the particular product. Nevertheless, please tick a circle in every line. \\
    \\
    It is your personal opinion that counts. Please remember: there is no wrong or right answer!\\
    \\
    \textbf{Please assess the model now by ticking one circle per line}\\

        \begin{tabular}{l|ccccccc|l}
        \hline
         & 1 & 2 & 3 & 4 & 5 & 6 & 7 & \\
        \hline
        obstructive & \Square & \Square & \Square & \Square & \Square & \Square & \Square & supportive\\
        complicated & \Square & \Square & \Square & \Square & \Square & \Square & \Square & easy\\
        inefficient & \Square & \Square & \Square & \Square & \Square & \Square & \Square & efficient\\
        confusing & \Square & \Square & \Square & \Square & \Square & \Square & \Square & clear\\
        boring & \Square & \Square & \Square & \Square & \Square & \Square & \Square & exciting\\
        Not interesting & \Square & \Square & \Square & \Square & \Square & \Square & \Square & interesting\\
        conventional & \Square & \Square & \Square & \Square & \Square & \Square & \Square & inventive\\
        usual & \Square & \Square & \Square & \Square & \Square & \Square & \Square & Leading edge\\
        \hline
        \end{tabular}

\end{table*}

\section{Quantitative Results}

\begin{table*}[hbt]
    \centering
    \caption{Overview of the Average Accuracy of Two Visualizations: average accuracy rates across three different stages — pre-test, immediate testing, and long-term testing, as well as three levels of difficulty (easy, moderate, and difficult) in Year and Location information.}
    \label{total accuracy}
   \renewcommand{\arraystretch}{1.1}
    \begin{tabular}{llccc}
        \toprule
        &\multicolumn{1}{c}{}& \multirow{2}{*}{\textbf{Pre-Test}} & \textbf{Immediate Memory} & \textbf{Memory} \\
        && ~ & \textbf{Comprehension} & \textbf{Retention} \\
        \midrule
        \multirow{4}{*}{Physicalization} & \multirow{2}{*}{Overall} & 35.16\% & 84.90\% & 73.96\% \\
        ~ & ~ & 46.88\%\textbar32.03\%\textbar26.56\% & 92.19\%\textbar83.59\%\textbar  78.13\% & 88.28\%\textbar78.13\%\textbar55.47\%  \\ 
        \cdashline{2-5}[0.8pt/2pt]
        ~ & Year & 20.83\% & 88.54\% & 76.04\% \\ 
        \cdashline{2-5}[0.8pt/2pt]
        ~ & Location & 49.48\% & 81.25\% & 71.88\% \\ 
        \midrule
        \multirow{4}{*}{Digitalization} & \multirow{2}{*}{Overall} & 31.25\% & 78.13\% & 58.85\% \\
        ~ & ~ & 37.50\%\textbar25.00\%\textbar31.25\% & 84.38\%\textbar74.22\%\textbar75.78\% & 66.93\%\textbar66.67\%\textbar40.36\%  \\ 
        \cdashline{2-5}[0.8pt/2pt]
        ~ & Year & 27.08\% & 84.38\% & 63.02\% \\ 
        \cdashline{2-5}[0.8pt/2pt]
        ~ & Location & 35.42\% & 71.88\% & 54.69\% \\ 
        \bottomrule  
    \end{tabular}
\end{table*}

\section{Qualatitive Result}

\begin{table*}[htbp]
    \centering
    \caption{Patterns described by participants using digital visualization}
    \label{pattern_digital}
    \renewcommand{\arraystretch}{1.1}
        \begin{tabular}{cp{.45\textwidth}p{.45\textwidth}}
        \toprule
        \textbf{ID} & \textbf{Immediate comprehension} & \textbf{Memory retention} \\
        \midrule
        \squarecolorbox{o}{ 1} & 1. Arrange the buildings from inside to outside according to the time they are put into use. & From the inside out in chronological order. \\
        ~ & 2. Buildings put into use in the same year are placed in the same circle.  & ~\\  \hdashline[0.8pt/2pt]
        
        \squarecolorbox{o}{ 2} & With FB as the center, the North campus is distributed in the north of FB, and the south campus is distributed in the south of FB. & With FB as the center, the North campus is on the north side and the south campus is on the south side. \\\hdashline[0.8pt/2pt]
        
        \squarecolorbox{o}{ 3} & The buildings are arranged chronologically from the inside out. & The models are arranged in layers of annual wheels according to the construction year of the school. \\\hdashline[0.8pt/2pt]
        
        \squarecolorbox{o}{ 4} & The circle represents the year equipotential line (from inside out indicates the service time from long to short). & 1. The closer the building is to the center, the longer it has been in use. \\
        ~ & ~ & \textcolor{emphasize}{2. The time gap between buildings being put into use can be judged by the distance between corresponding circles.}\\\hdashline[0.8pt/2pt]
        
        \squarecolorbox{o}{ 5} & From the center to the sides, the year is getting newer and newer. & The location of the building in the timeline is the same as in real life. \\\hdashline[0.8pt/2pt]
        
        \squarecolorbox{o}{ 6} & The buildings on the same circle are placed according to the actual geographical location. & Each model conforms to the east-west direction in reality. \\\hdashline[0.8pt/2pt]
        
        \squarecolorbox{o}{ 7} & 1. The innermost part of the disc represents the earliest year when it is put into use, and the years are evenly distributed. & It is logical and clear to memorize. \\
        ~ & \textcolor{emphasize}{2. The denser the disc is, the closer the time it is put into use.} & ~ \\ \hdashline[0.8pt/2pt]
        
        \squarecolorbox{o}{ 8} & Buildings put into use in the same year are on the same circle. & 1. Each building of the school is in a time circle with the earliest time in the inner circle and the latest time in the outer circle.\\ 
         ~ & ~ & 2. The location of the building in the timeline is the same as in real life.\\\hdashline[0.8pt/2pt]
         
         \squarecolorbox{o}{ 9} & \textcolor{emphasize}{1. The North campus to be put into use first and the South campus to be put into use later.} & 1. The building is built from the inside out, and the year when the building is put into use increases. \\
        ~ & 2. Put into use the building with higher columns first, and then put into the shorter building & \textcolor{emphasize}{2. Fewer buildings are put into use in the early stage and more intensive in the later stage.}\\
        ~ & ~ & \textcolor{emphasize}{3. The buildings on the North campus are put into use first, followed by those on the South campus.}\\\hdashline[0.8pt/2pt]
        
        \squarecolorbox{o}{ 10} & 1. The closer to the centre, the earlier the building is put into use. & 1. The oldest building located at the centre and the highest. The newest building located away from the centre. \\
        ~ & 2. Buildings put into the same year are located on the same circle. & 2. Buildings put into use in the same year are on the same annual rings. \\   
        ~ & 3. The distance of gaps between circles are correlated to the years. & ~ \\\hdashline[0.8pt/2pt]
        
        \squarecolorbox{o}{ 11} & The building is from the inner circle to the outer circle, from high to low, representing the year of the building being put into use from early to late. & 1. The year when the building was put into use is very clear, and the buildings on the same horizontal line can be remembered by association.\\ 
        ~ & ~ & 2. The height of the buildings in different years is different, so it is easy to remember the order of the years. \\\hdashline[0.8pt/2pt]
        
        \squarecolorbox{o}{ 12} & The year marked on the bottom disk tells you the year each building is put into use. & This model presents the various buildings in the north and south campuses on a concentric circle with time as the axis, based on their geographical location and relative distance. \\\hdashline[0.8pt/2pt]
        
        \squarecolorbox{o}{ 13} & 1. The buildings of the South Campus are distributed symmetrically. & 1. This is a circular model that integrates the location of the building and the construction time. \\
         ~ & \textcolor{emphasize}{2. The use time of the South Campus is shorter when the overall building is put into use, while the North Campus is longer when the overall building is put into use.} & 2. The farther away from the center point of the circular disk, the larger the radius, and the shorter the construction time. \\\hdashline[0.8pt/2pt]
        
        \squarecolorbox{o}{ 14} & \textcolor{emphasize}{The closer the buildings are, the more similar the years they are used.} & \textcolor{emphasize}{1. The closer the building is, the closer the year it was put into use.}\\ 
        ~ & ~ & 2. The layout of the building corresponds to the actual situation. \\\hdashline[0.8pt/2pt]
        
        \squarecolorbox{o}{ 15} & 1. All the buildings are allocated according to both time lines and their actual locations. & 1. The model adopts the form of time-space distribution, and displays the building model from the inside out according to the time of building construction in the horizontal direction. \\
         ~ & \textcolor{emphasize}{2. Additionally, the general trend shows that all the buildings in the north campus appear earlier than those in south.} & 2. Since the buildings in different time loops are arranged according to the actual orientation, I can intuitively and quickly understand what the model wants to describe. \\\hdashline[0.8pt/2pt]
        
        \squarecolorbox{o}{ 16} & The closer it is to the center of the circle, the earlier the building will be put into use. & 1. The farther from the center of the circle, the newer the year the building was put into use.\\ 
        ~ & ~ & 2. The color of the building model is consistent with the reality. \\ 
        \bottomrule
        \end{tabular}
\end{table*}

\begin{table*}[htbp]
    \centering
    \caption{Patterns described by participants using physical visualization}
    \label{pattern_physical}
    \renewcommand{\arraystretch}{1.1}
        \begin{tabular}{cp{.45\textwidth}p{.45\textwidth}}
        \toprule
        \textbf{ID} & \textbf{Immediate comprehension} & \textbf{Memory retention} \\
        \midrule
        \squarecolorbox{b}{ 17} & The buildings are distributed according to time, and the inner circle buildings are put into use earliest. & The building gradually transforms from inside to outside. \\\hdashline[0.8pt/2pt]                                                                                                                             
        \squarecolorbox{b}{ 18} & \textcolor{emphasize}{1. The buildings of the South Campus are put into use from 2016 to 2018.} & 1. The model distributed every building in XJTLU in a circle, with the latest outside and the oldest inside. \\
        ~ & \textcolor{emphasize}{2. The buildings of the North Campus are all put into use before 2013.} & 2. The buildings located from west to the east is allocated from the left to the right in the model. \\
        ~ & \textcolor{emphasize}{3. The North campus buildings are put into use in a relatively scattered time, while the South campus buildings are more concentrated.}& ~ \\
        ~ & 4. The buildings placed in the outer ring are newer.& ~ \\\hdashline[0.8pt/2pt] 
        
        \squarecolorbox{b}{ 19} & Buildings are prioritized according to time, and buildings of the same year refer to the actual geographical location. & Use established years and locations to distinguish different buildings, and the established years are shown in concentric circles. \\\hdashline[0.8pt/2pt]
       
        \squarecolorbox{b}{ 20} & Building height gradually decreases from inside to outside. & Building height gradually decreases from inside to outside. \\

        &&\\\hdashline[0.8pt/2pt]
        
        \squarecolorbox{b}{ 21} & Buildings in the same circle are distributed according to their actual geographical location. & 1. The buildings near the center of the circle were put into use the earliest.\\ 
        ~ & ~ & 2. The distribution is roughly based on the actual geographical location. \\\hdashline[0.8pt/2pt]
        
        \squarecolorbox{b}{ 22} & \textcolor{emphasize}{1. The buildings of the North Campus are generally put into use earlier than those of the South Campus.} & \textcolor{emphasize}{The buildings of the North Campus were put into use earlier than those of the South Campus.} \\ 
        ~ & \textcolor{emphasize}{2. The building of the South Campus has been put into use for a shorter time than that of the North Campus.}& ~ \\\hdashline[0.8pt/2pt]
        
        \squarecolorbox{b}{ 23} & \textcolor{emphasize}{1. The buildings of the North Campus are generally put into use earlier.} & Buildings near the center of the circle are invested the earliest. \\ 
        ~ & \textcolor{emphasize}{2. The construction of the South Campus is generally late and time intensive.} & ~ \\\hdashline[0.8pt/2pt]
        
        \squarecolorbox{b}{ 24} & The different heights of the model buildings indicate the time of being put into use. & 1. The position of the model is designed in concentric circles, which makes it easy to remember which buildings are in the same year.\\ 
        ~ & ~ & 2. Different models have different color heights and appearance, giving people a deeper impression. \\\hdashline[0.8pt/2pt]
        
        \squarecolorbox{b}{ 25} & All the buildings outside are put into use relatively recently. & The closer the building is placed to the outer ring, the newer the year it is put into use. \\\hdashline[0.8pt/2pt]
        
        \squarecolorbox{b}{ 26} & \textcolor{emphasize}{The number of buildings put into use in the early stage is small, and the number of buildings put into use in the later stage is more.} & This 3D model contains the content of the year and geography information and so on. \\\hdashline[0.8pt/2pt]
        
        \squarecolorbox{b}{ 27} & \textcolor{emphasize}{The buildings of North campus are put into use earlier, while those of South Campus are put into use later.} & It is a 3D model. \\\hdashline[0.8pt/2pt]
        
        \squarecolorbox{b}{ 28} & \textcolor{emphasize}{The closer the buildings are to the time when they are put into use, the closer the models are.} & 1. Building investment time gradually increases from inside to outside.\\
        ~ & ~ & \textcolor{emphasize}{2. The closer the time the building is put into use, the more compact it is.}\\\hdashline[0.8pt/2pt]
        
        \squarecolorbox{b}{ 29} & \textcolor{emphasize}{1. More buildings have been put into use in the South Campus.} & \textcolor{emphasize}{The South Campus building was put into use relatively late.} \\
        ~ & \textcolor{emphasize}{2. The buildings of the North Campus are put into use relatively early.} & ~ \\\hdashline[0.8pt/2pt]
        
        \squarecolorbox{b}{ 30} & \textcolor{emphasize}{1. The buildings of South Campus are put into use in 2016-2018.} & 1. The pattern is combined in time and space. \\
        ~ & \textcolor{emphasize}{2. The buildings on the south campus are relatively concentrated, while the buildings on the North campus are relatively evacuated.} & 2. But it is only easier for me to remember the relative location instead of the specific location.\\\hdashline[0.8pt/2pt]
        
        \squarecolorbox{b}{ 31} & 1. The location of the South campus buildings are near the periphery of the ring (the year they are put into use is relatively new). & 1. It's a circular model. \\ 
        ~ & 2. The North campus buildings are near the circle center. & 2. The closer to the center, the older the building.\\
        ~ & ~ & 3. The farther away from the center, the closer to the present time it was built. \\\hdashline[0.8pt/2pt]
        
        \squarecolorbox{b}{ 32} & 1. The year in which the building is put into use gradually decreases from the inside out. & 1. Circular distribution.\\ 
        ~ & 2. The building is centered on FB, which is the first put into use. & 2. The closer to the center of the circle, the older the building is. \\
        \bottomrule
        \end{tabular}
\end{table*}

Overall, we can see that users' understanding of patterns and insights may not align with those trained in data visualization, who typically explore from a data perspective. General users may instead derive patterns based on the visualization encoding rules. Therefore, we will provide further guidance on pattern and insight identification in future experiments to obtain more reasonable and accurate insight discoveries.

\end{document}